\documentclass[10pt,twocolumn,prl,aps,amssymb,amsmath,tightenlines,showpacs]{revtex4}
\usepackage{dsfont}
\usepackage{graphicx}
\usepackage{amssymb,amsfonts}
\usepackage{overpic}
\usepackage[bb=boondox]{mathalfa}

\newcommand{\half}{\mbox{$\textstyle \frac{1}{2}$}}
\newcommand{\re}{\mbox{$\rm e$}}
\newcommand{\ri}{\mbox{$\rm i$}}
\newcommand{\rd}{\mbox{$\rm d$}}

\begin{document}

\title{Time-optimal navigation through quantum wind}

\author{Dorje C. Brody${}^{1,2}$, Gary W. Gibbons${}^{3}$, and David M. Meier${}^{1}$}

\affiliation{${}^{1}$Department of Mathematics, Brunel University London, Uxbridge UB8 3PH, UK \\ 
${}^{2}$St Petersburg National Research University of Information Technologies, Mechanics and Optics,\\ 
Kronwerkskii ave 49,  St Petersburg, 197101, Russia \\
${}^{3}$Department of Applied Mathematics and Theoretical Physics, Centre for Mathematical 
Sciences, Wilberforce Road, Cambridge CB3 0WA, UK } 

\date{\today}

\begin{abstract}
\noindent 
The quantum navigation problem of finding the time-optimal control Hamiltonian that transports 
a given initial state to a target state through quantum wind, that is, under the influence of external 
fields or potentials, is analysed. By lifting the problem from the state space to the space of unitary gates 
realising the required task, we are able to deduce the form of the solution to the problem by 
deriving a universal quantum speed limit. The expression thus obtained indicates that further 
simplifications of this apparently difficult problem are possible if we switch to the interaction 
picture of quantum mechanics. A complete solution to the navigation problem for an arbitrary 
quantum system is then obtained, and the behaviour of the solution is illustrated in the case of 
a two-level system. 
\end{abstract}


\maketitle

\noindent 
With the advances in the implementation of quantum technologies, the theoretical understanding of 
controlled quantum dynamics and, in particular, of their limits, is becoming increasingly important. 
One aspect of such limits that has been investigated extensively in the literature concerns the 
time-optimal manoeuvring of quantum states \cite{ML,lloyd1,brockett,kosloff,brody1,lloyd3,SSKG,
hosoya,BM,brody2,hosoya2,zanardi,caneva,bloch,GCH,garon,stepney1,lloyd}. If the time-evolution 
is unconstrained (apart from a 
bound on the energy resource), then this amounts to finding the time-independent Hamiltonian 
that generates maximum speed of evolution. 
However, in general there can be a range of 
constraints that prohibits the implementation of such an elementary protocol, and various 
optimisations will have to be applied to determine time-dependent Hamiltonians that generate 
the dynamics achieving required tasks.  

An important class of problems arising in this context is the identification of the time-optimal 
quantum evolution under the influence of external fields or potentials that cannot be easily 
eliminated in a laboratory. Solutions to such problems are indeed relevant to practical 
implementations of time-optimal controlled quantum dynamics because in real laboratories 
external influences (e.g, electromagnetic fields) are typically present. Problems of this kind 
can be thought of as representing the quantum counterpart of the classical Zermelo navigation 
problem of finding the time-optimal control that takes a ship from one location to another, under 
the influence of external wind or currents \cite{zermelo,caratheodory}. 
Within the context of quantum Zermelo problems, 
there are two distinct questions that arise, namely, 
finding the time-optimal Hamiltonian (i) that generates a required unitary gate; and (ii) that 
transports a given initial state to a required target state. In the context of constructing a unitary 
gate, this problem was formulated in \cite{stepney2}, and solved more recently in 
\cite{stepney3,brody3}. 
The construction of the time-optimal Hamiltonian that transports a given initial state to a target 
state, on the other hand, is \textit{a priori} more challenging (cf. \cite{BM1}), because 
(as shown below) this involves a variational problem with free boundaries. The purpose of the 
present paper is to derive the full solution to this latter problem for an arbitrary quantum system. 

One way of addressing the quantum-state navigation problem is to work in the space of rays 
through the origin of the underlying Hilbert space (i.e. the complex projective space, 
or simply 
the `state space'). Then the 
Schr\"odinger evolution generated by the background field gives rise to a Hamiltonian vector 
field on the 
state space, and the task at hand can be formulated as a Hamiltonian control problem, 
the solution of which is typically difficult to obtain. We are nevertheless able to go forward by 
exploiting certain subtle geometric structures of the unitary group, which in turn allows us to 
efficiently make use of the results obtained in \cite{stepney3,brody3}. 
A remarkable feature of the solution that we obtain is that it is strikingly reminiscent of the 
analysis of quantum dynamics in the interaction-picture of Schwinger and Tomonaga 
\cite{sunakawa}. 
On account of this observation we are able  to revisit our analysis by employing a 
physical symmetry argument, rather than purely variational reasoning. As a consequence, 
a considerably simpler derivation of the solution emerges. In fact, it turns out that by invoking 
the line of thinking behind the Schwinger-Tomonaga theory, a straightforward solution can 
even 
be found of closely-related and apparently difficult problems in geometry \cite{robles0,robles}. 
For an illustration 
we conclude the paper by presenting examples in the case of a two-level system. 

Let us begin by stating more explicitly the Zermelo navigation problem for 
quantum states: For a given initial state $|\psi_I\rangle$ and a final target state $|\psi_F\rangle$, 
the task of an experimentalist is to find the control Hamiltonian ${\hat H}_1(t)$ such that the total 
Hamiltonian ${\hat H}(t)={\hat H}_0+{\hat H}_1(t)$ will generate the transformation $|\psi_I\rangle 
\to |\psi_F \rangle$ in the shortest possible time. Here ${\hat H}_0$ represents the Hamiltonian 
of the background field or potential that cannot be manipulated. Since the objective is to 
realise a time-optimal transformation (navigation), it is assumed that the available energy 
resource, which evidently has to be bounded, will be consumed fully. This translates into 
the `full throttle' condition that the squared magnitude of the evolution speed generated by the 
control Hamiltonian ${\hat H}_1$, which is given on account of the Anandan-Aharonov relation 
\cite{AA} by four times the variance $4\Delta{H}_1^2$ of ${\hat H}_1$, is held fixed at all times 
at the maximum attainable value. 

\begin{figure}[t]
\begin{center}
\includegraphics[scale=0.45]{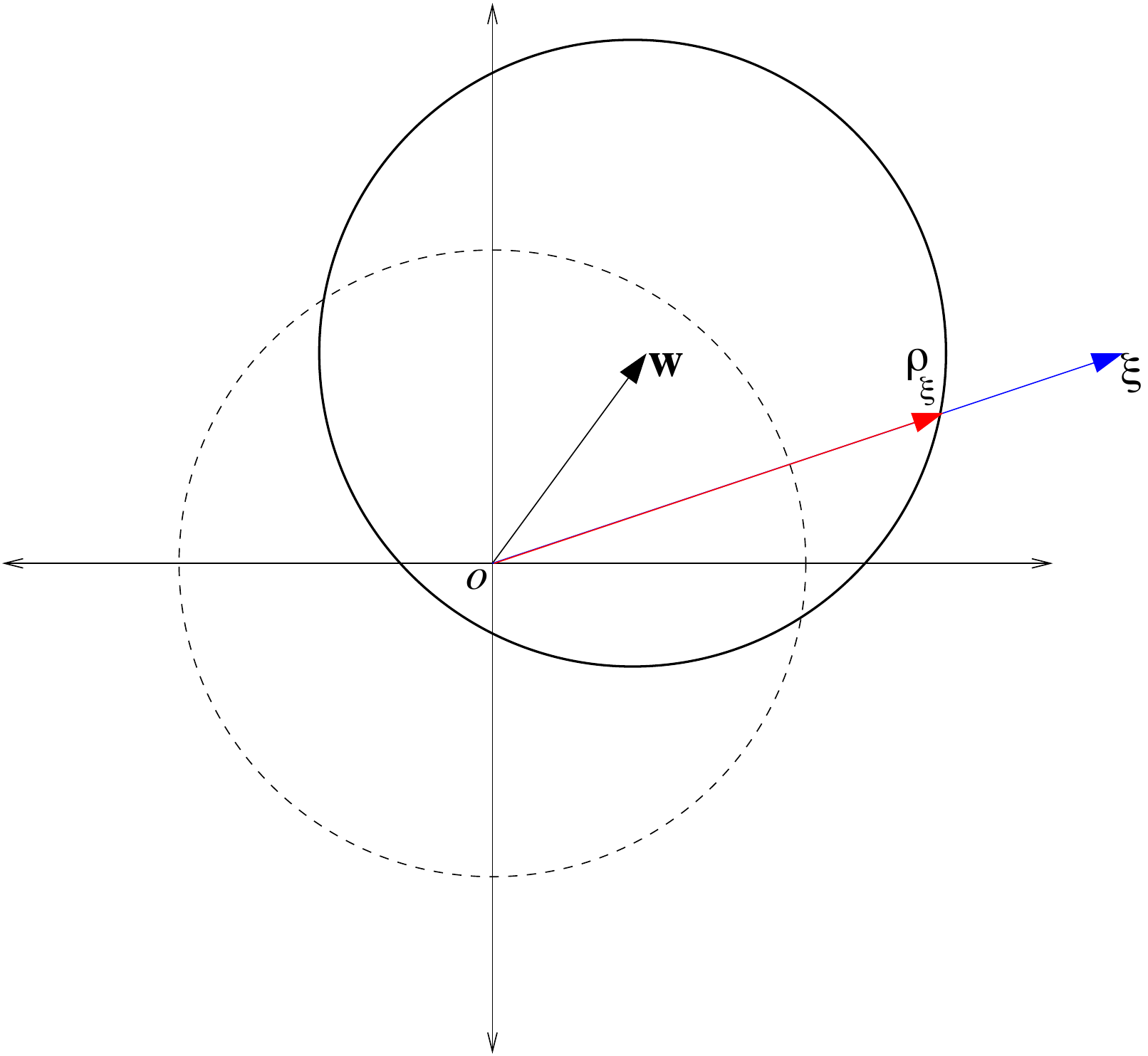}
\caption{
\label{fig:1}
(colour online) 
\textit{Unit circle in the wind}. In the absence of the wind, the `unit-time' circle is 
centred at the origin $o$. However, with the prevailing wind, with strength 
$|\vec{w}|<1$, 
the unit-time circle reachable from the origin is shifted in the direction of the wind. Since 
it takes one unit of time to reach $\rho_\xi$, reaching the target $\xi$ will take 
$|\protect\overrightarrow{o\xi}| / |\protect\overrightarrow{o\rho}_\xi|$ units of time.}
\end{center}
\end{figure}

To illustrate the nature of the task involved, let us first consider a more elementary problem 
of Zermelo navigation on the plane ${\mathds R}^2$, with constant (in space 
and in time) wind, represented by the vector $\vec{w}$. Suppose that, departing from the origin 
$o$ the desired destination is given by the endpoint 
of a vector $\vec{\xi}$. To work out the time it takes to reach the destination, consider a circle 
whose radius is determined by the distance that can be reached with full speed over one unit of 
time. In the absence of wind the circle is clearly centred at the origin. However, under the influence 
of the wind the centre of the circle is shifted by $\vec{w}$. For the moment let us assume that 
the wind is not dominant, i.e. $|\vec{w}|<1$ so that the origin remains inside the shifted circle. 
This configuration is schematically illustrated in figure~\ref{fig:1}. Under the influence of the wind, 
therefore, to reach the endpoint $\xi$ of the vector $\vec{\xi}$, it suffices to determine the distance 
$|\overrightarrow{o\rho_\xi}|$, where $\rho_\xi$ is the point of intersection of the vector 
$\overrightarrow{o\xi}$ with the shifted circle. This follows on account of the fact that since it 
takes one unit of time to reach $\rho_\xi$, it will take $F(\xi)=|\overrightarrow{o\xi}| / 
|\overrightarrow{o\rho}_\xi|$ units of time to reach $\xi$. Elementary algebra then shows that 
the required time $F(\xi)$ is given by 
\begin{eqnarray}
F(\xi) = \frac{\sqrt{\langle \vec{w},\vec{\xi}\rangle^2+|\vec{\xi}|^2(1-|\vec{w}|^2)} 
- \langle \vec{w},\vec{\xi}\rangle}{1-|\vec{w}|^2}, 
\label{eq:1}
\end{eqnarray}
where $\langle \vec{w},\vec{\xi}\rangle=\vec{w}\cdot\vec{\xi}$ is the Euclidean inner product,  
$|\vec{w}|^2=\langle \vec{w},\vec{w}\rangle$, and $|\vec{\xi}|^2=\langle \vec{\xi},\vec{\xi}\rangle$. 

The foregoing analysis on ${\mathds R}^2$ extends naturally to the general case of navigation 
on a Riemannian manifold of arbitrary dimension, albeit the Euclidean inner product 
$\langle~,~\rangle$ has to be replaced by the Riemannian one. This is because the total journey 
time can be decomposed into a large number of infinitesimal journey 
times, each on a tangent space of the manifold. In the case of optimal quantum navigation, 
we recall that the space of pure states is equipped with the unitary-invariant Fubini-Study metric 
\cite{kibble,gibbons,brody0}. 
The wind $\vec{w}$ is then replaced by the Hamiltonian symplectic flow generated by ${\hat H}_0$, 
whereas $\vec{\xi}$ is some vector tangent to quantum state space.  
The total journey time $T$ along a path on state space is given by the integral of $F(\xi)$, where 
$\vec{\xi}$ is the velocity vector. 
In particular, for the Zermelo problem, $\vec{\xi}$ is generated by 
${\hat H}(t)={\hat H}_0+{\hat H}_1(t)$, and this implies that $F(\xi)$ in the quantum context is given by a 
function of the variance $\Delta{H}_0^2$ of ${\hat H}_0$, the variance $\Delta{H}_1^2$ of 
${\hat H}_1$, and the covariance $\langle{\hat H}_0{\hat H}_1\rangle$ of the two Hamiltonians.  
The optimal control is obtained by minimising $T$ over all ${\hat H}_1(t)$, 
subject to the full throttle constraint $4\Delta{H}_1^2=1$. The no-dominance condition on the 
wind is required in the case of a navigation 
problem on an open manifold such as ${\mathds R}^n$, since otherwise the target may never 
be reached. In the quantum context we are able to relax the requirement 
$4\Delta{H}_0^2<1$, since the manifold 
of pure states is compact, which means that if the wind against the shortest path is too strong, 
one can always go `the other way around' to reach the target state (note incidentally that 
$F(\xi)$ remains well defined in the limit $|\vec{w}|\to1$ since 
$F(\xi)\to|\vec{\xi}|^2/2\langle \vec{w},\vec{\xi}\rangle$). 

The Hamiltonian control problem specified above is not straightforward to solve directly, 
and this leads us to employ an alternative approach based on the idea of lifting the problem 
from state space to the unitary group acting on the states. In fact, from a physical 
point of view this is more appealing, since, although we are seeking the time-optimal path on 
state space, physical implementations are always carried out by constructing the Hamiltonian 
that generates the desired unitary operator. 

\begin{figure}[t]
\begin{center}
\begin{overpic}[scale = 0.0575]{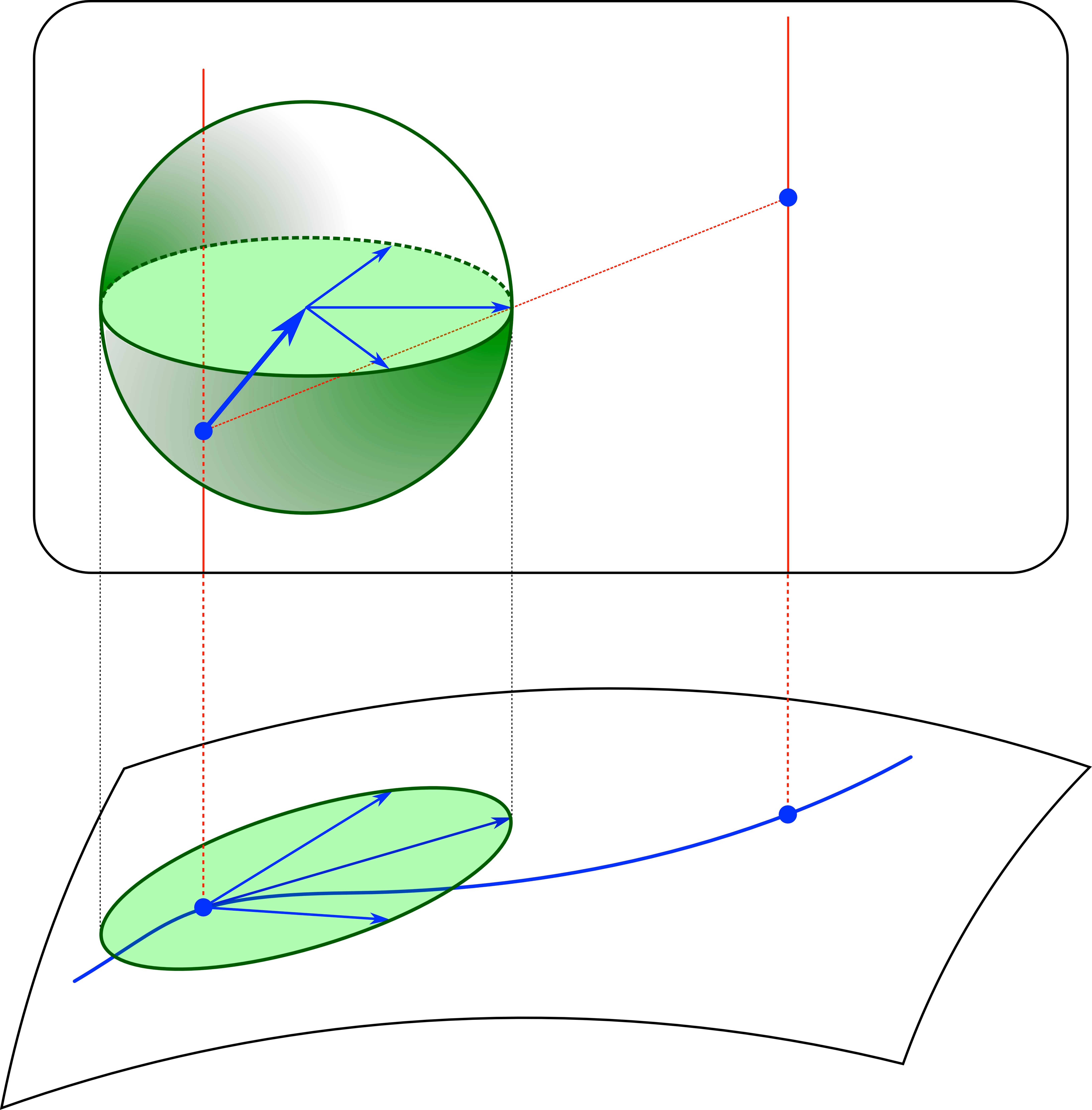}
\put(72.5, 79.5){${\hat u}_F$}
\put(16.40, 14) {$|\psi_I\rangle$}
\put(19.50, 58) {${\hat e}$}
\put(22.50, 73) {${\hat H}_0$}
\put(68.50, 22) {$|\psi_F\rangle$}
\put(74.0, 9.0) {${\mathds C}{\mathbb P}^n$}
\put(76.0, 52) {${\rm SU}(n+1)$}
\put(74.5, 41.50) {$\left\Downarrow~{\hat u}\to{\hat u}|\psi_I\rangle\right.$}
\end{overpic}
\end{center}
\caption{
\label{fig:2}
(colour online) 
\textit{Lifting of the navigation problem}. We construct a fibre space of unitary group 
elements 
above the state space with the property that any element ${\hat u}$ in the fibre above the 
state $|\psi\rangle$ is projected down according to ${\hat u}|\psi_I\rangle=|\psi\rangle$. 
Starting from the identity element ${\hat e}\in{\rm SU}(n+1)$ in the fibre above 
$|\psi_I\rangle$, the unitary group elements that can be reached in one unit of time 
under the influence of the `wind' ${\hat H}_0$ form a shifted unit sphere. The projection 
of the sphere onto the state space ${\mathds C}{\mathbb P}^n$ then determines the 
set of states reachable in that time. The furthest that one can transport the state is by 
moving away from ${\hat e}$ in the horizontal direction. There remain many 
horizontal directions, one of which is singled out by the boundary condition 
${\hat u}_F|\psi_I\rangle=|\psi_F\rangle$.}
\end{figure}
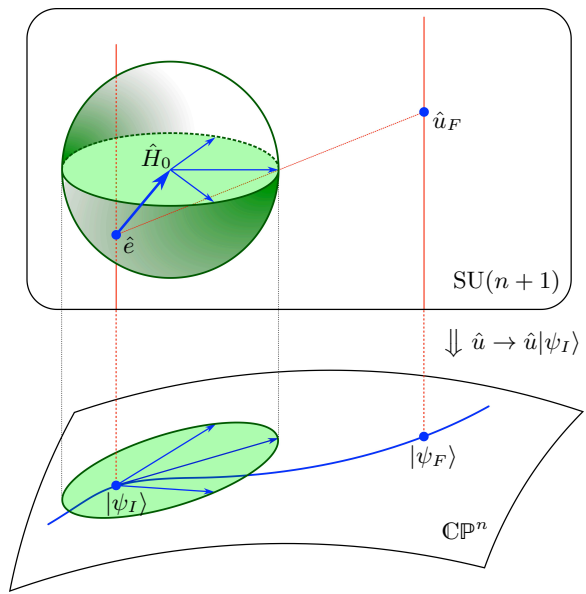

With this in mind, we proceed by considering a fibre space above the state space with the 
property that for any 
state $|\psi\rangle$, the fibre above it consists of elements $\{{\hat u}\}$ of the special unitary 
group fulfilling the condition that ${\hat u}|\psi_I\rangle=|\psi\rangle$. Although our solution is 
applicable to both finite and infinite dimensional Hilbert spaces, to simplify the discussion let 
us assume that the Hilbert space is of finite dimension $n+1$. The totality of such fibres, 
when bundled together, then forms the group SU$(n+1)$, which acts on the state space 
(the complex projective space 
${\mathds C}{\mathbb P}^n$). This configuration is illustrated in figure~\ref{fig:2}. Let 
${\hat u}_F$ be an element of the fibre above the target state $|\psi_F\rangle$, i.e. 
${\hat u}_F|\psi_I\rangle=|\psi_F\rangle$. Then the problem of finding the time-optimal 
transformation $|\psi_I\rangle\to|\psi_F\rangle$ translates into the problem of constructing 
the unitary gate ${\hat u}_F$ from the identity element ${\hat e}\in{\rm SU}(n+1)$ in the 
shortest possible time, under the influence of an external field. This 
translation of the problem helps us because the optimal Zermelo navigation for a specified 
unitary gate has been worked out recently \cite{stepney3,brody3}, and we can make use of 
the result in deducing the optimal control Hamiltonian for the state transfer; however, we 
encounter two difficulties: (a) the `target' unitary gate ${\hat u}_F$ is not 
unique since \textit{a priori} it can be any element of the fibre above $|\psi_F\rangle$. In other 
words, if ${\hat v}_F$ is any unitary operator that leaves the state $|\psi_F\rangle$ invariant, 
then the unitary gate ${\hat v}_F{\hat u}_F$ also satisfies the required condition ${\hat v}_F
{\hat u}_F |\psi_I\rangle=|\psi_F\rangle$ so we have to deal with a free final boundary; 
(b) the `full throttle' condition $4\Delta{H}_1^2=1$ on 
state space is concerned with the evolution speed, whereas the condition used in 
\cite{stepney2,stepney3,brody3} for unitary gates appears to be distinct since it 
concerns the Hilbert-Schmidt norm $2{\rm tr}({\hat H}_1^2)=1$. In what follows we shall show that 
these two apparent issues evaporate once the relevant `horizontality condition' is imposed. 

To see this we begin by noting that since ${\hat e}|\psi_I\rangle=
|\psi_I\rangle$, the identity element ${\hat e}$ belongs to the fibre above the initial state 
$|\psi_I\rangle$. In fact, the fibre above $|\psi_I\rangle$ coincides with the totality of unitary 
group elements $\{{\hat v}_I\}$ that leave the initial state $|\psi_I\rangle$ invariant. Intuitively, 
to reach from ${\hat e}$ a target gate ${\hat u}_F$ above $|\psi_F\rangle$ in a timely manner 
we would like to move away from ${\hat e}$ as quickly as possible, and this is achieved by 
manoeuvring \textit{horizontally}, i.e. towards a direction that is orthogonal to the fibre. 
That is to say, the choice of the 
initial control Hamiltonian ${\hat H}_1(0)$ has to be such that ${\rm tr}({\hat H}_1(0) {\hat H}_I)
=0$ for all Hamiltonians ${\hat H}_I$ that leave the initial state invariant (i.e. ${\hat H}_1(0)$ 
has to be orthogonal to all generators of ${\hat v}_I$). This suggests that the horizontality 
condition ${\rm tr}({\hat H}_1(0) {\hat H}_I)=0$
gives rise to the maximum evolution speed. To show this more explicitly we shall derive a kind 
of universal 
quantum speed limit. The statement of the result we shall establish is as follows: The 
squared speed of the evolution of a quantum state generated by a Hamiltonian ${\hat H}$, as 
defined by Anandan and Aharonov \cite{AA}, is bounded above by twice the Hilbert-Schmidt norm 
${\rm tr}({\hat H}^2)$ of the Hamiltonian, and the bound is attained if ${\hat H}$ is horizontal. 
The implication of this result 
is that under the horizontality requirement the norm condition and the maximum speed 
condition are equivalent, and this resolves the issue (b) raised above. 
As a consequence, we are able to borrow the results of 
\cite{stepney3,brody3} to 
deduce that the time-optimal control Hamiltonian ${\hat H}_1(t)$ 
is necessarily of the form 
\begin{eqnarray}
{\hat H}_1(t) = \re^{-{\rm i}{\hat H}_0 t} {\hat H}_1(0) \re^{{\rm i}{\hat H}_0 t} . 
\label{eq:2} 
\end{eqnarray}
Notice that the horizontality condition is preserved under the adjoint action (\ref{eq:2}), 
which is indeed required since the full-throttle condition has to be maintained 
throughout the operation

The horizontality condition alone, of course, does not fix ${\hat H}_1(0)$ uniquely, since the 
constraint ${\rm tr}({\hat H}_1(0) {\hat H}_I)=0$ on ${\hat H}_1(0)$ leaves $2n$ degrees of 
freedom. However, by demanding that the required gate should generate the target 
state $|\psi_F\rangle$, which imposes $2n$ conditions, we are able to select a unique 
initial control Hamiltonian ${\hat H}_1(0)$. In other words, the freedom in the boundary is 
eliminated by the horizontality condition ${\rm tr}({\hat H}_1(0) {\hat H}_I)=0$ along with the 
boundary condition ${\hat u}_F|\psi_I\rangle=|\psi_F\rangle$, and this resolves the issue (a) 
raised above. 

Before we turn to determining the initial control Hamiltonian ${\hat H}_1(0)$, and the amount of 
time $T$ that is needed to reach the target state under the optimal control, let us discuss the 
claim above on the universal speed limit (cf. \cite{raam}). From a physical point of view the 
claim is plausible if we make note of the following observation. Suppose that the Hamiltonian 
contains a vertical component tangent to the fibre that leaves the state below invariant. In this 
case, although energy is scarce, 
the experimentalist will be consuming energy that produces no work (except for shifting the 
overall phase). This is clearly not energy efficient. It follows that the optimal performance is ensured 
by eliminating vertical components altogether. A more precise derivation, which essentially follows 
from a standard result on complex projective spaces outlined, e.g., in Kobayashi \& Nomizu 
\cite{K-N} (see \S~XI.10), is given in the Supplementary Information. 

Having established the `geodesic' curve (\ref{eq:2}) that minimises the action $T=\int F \rd s$ 
we are now in the position to identify the initial control ${\hat H}_1(0)$ so that the target state 
is reached. To this end we observe that the form (\ref{eq:2}) of the solution is indicative of the 
analysis in the interaction picture of quantum mechanics, if we view ${\hat H}_0$ as the `free' 
and ${\hat H}_1(0)$ as the `interaction' Hamiltonian. Alternatively stated, if we change from the 
`rest frame' to the `moving frame' on the state space according to the wind ${\hat H}_0$, then 
without the control Hamiltonian the initial state remains fixed, while the target state moves against 
the wind according to $|\psi_F(t)\rangle = \re^{{\rm i}{\hat H}_0t}|\psi_F\rangle$. Evidently, the 
state $|\psi_F(t)\rangle$ thus defined is the solution to the (nonrelativistic version of) the 
Tomonaga-Schwinger equation. In particular, standard results in the interaction-picture analysis 
\cite{sunakawa} show that the time-ordered product of the unitary group generated by the 
Hamiltonian 
${\hat H}(t) = {\hat H}_0 + \re^{-{\rm i}{\hat H}_0 t} {\hat H}_1(0) \re^{{\rm i}{\hat H}_0 t}$ gives 
rise to the evolution equation of the form 
\begin{eqnarray}
|\psi(t)\rangle = \re^{-{\rm i}{\hat H}_0 t} \re^{-{\rm i}{\hat H}_1(0) t} |\psi_I\rangle . 
\label{eq:3}
\end{eqnarray} 
This result can also be verified directly (cf. \cite{stepney3}) by differentiation: writing 
${\hat u}(t)=\re^{-{\rm i}{\hat H}_0 t} \re^{-{\rm i}{\hat H}_1(0) t}$, we find $\partial_t{\hat u}(t) 
= -\ri {\hat H}(t) {\hat u}(t)$. As indicated above, to fix the initial condition ${\hat H}_1(0)$ we 
need to make use of the boundary condition, which implies that 
\begin{eqnarray}
\re^{-{\rm i}{\hat H}_0 T} \re^{-{\rm i}{\hat H}_1(0) T} |\psi_I\rangle = |\psi_F\rangle 
\label{eq:4}
\end{eqnarray}
must hold for some $T$ that is to be determined. Observe that there are $2n$ unknowns in 
${\hat H}_1(0)$, and we have an additional unknown parameter $T$. The boundary condition 
(\ref{eq:4}) gives rise to $2n$ constraints. Together with the norm condition 
$2{\rm tr}({\hat H}_1^2)=1$ we are thus able to fix all the unknowns, and this in turn solves the 
quantum-state Zermelo navigation problem. 

\begin{figure}[t]
\begin{center}
\begin{overpic}[scale = 0.205]{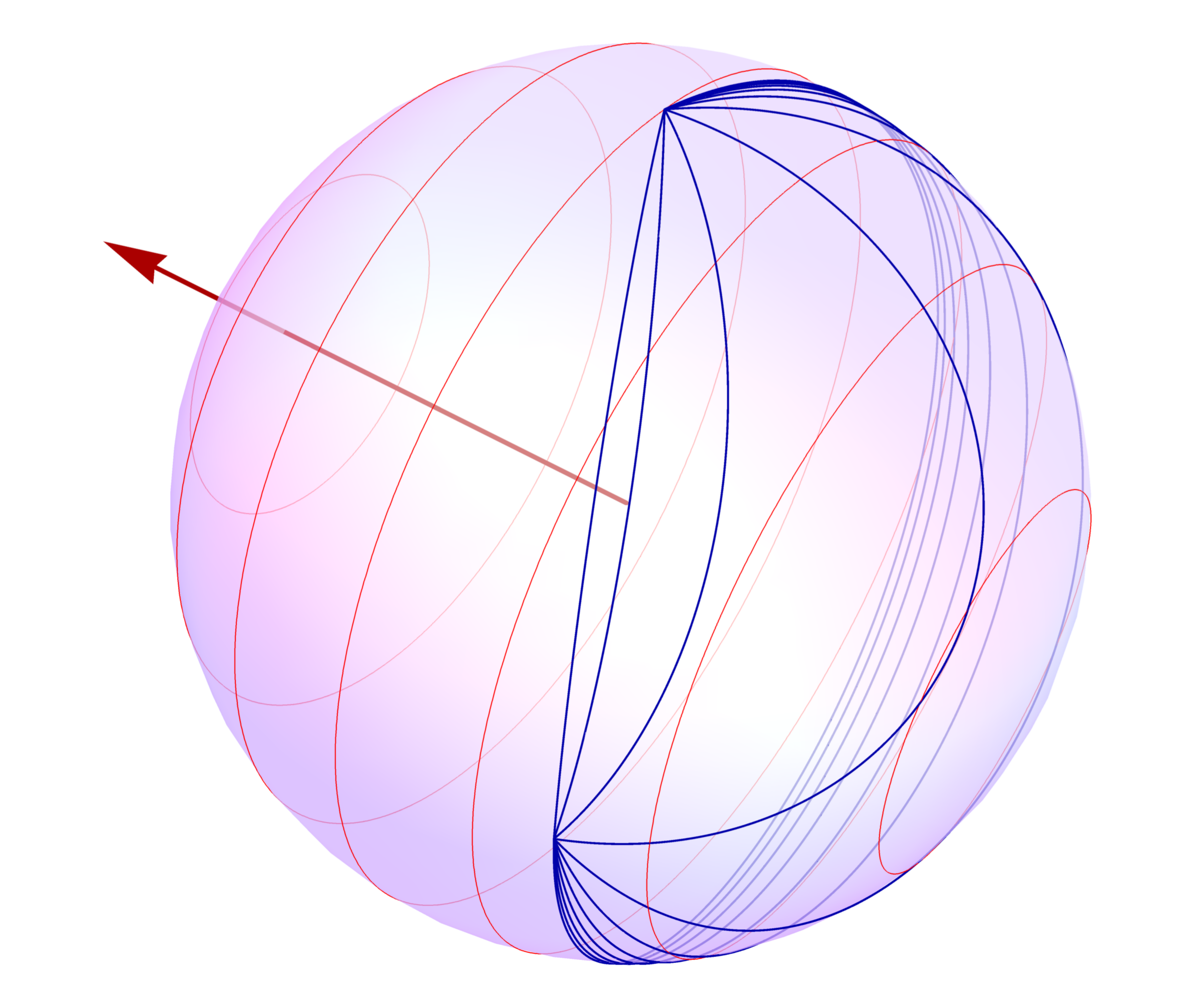}
\put(3, 54){\includegraphics[scale = 0.05]{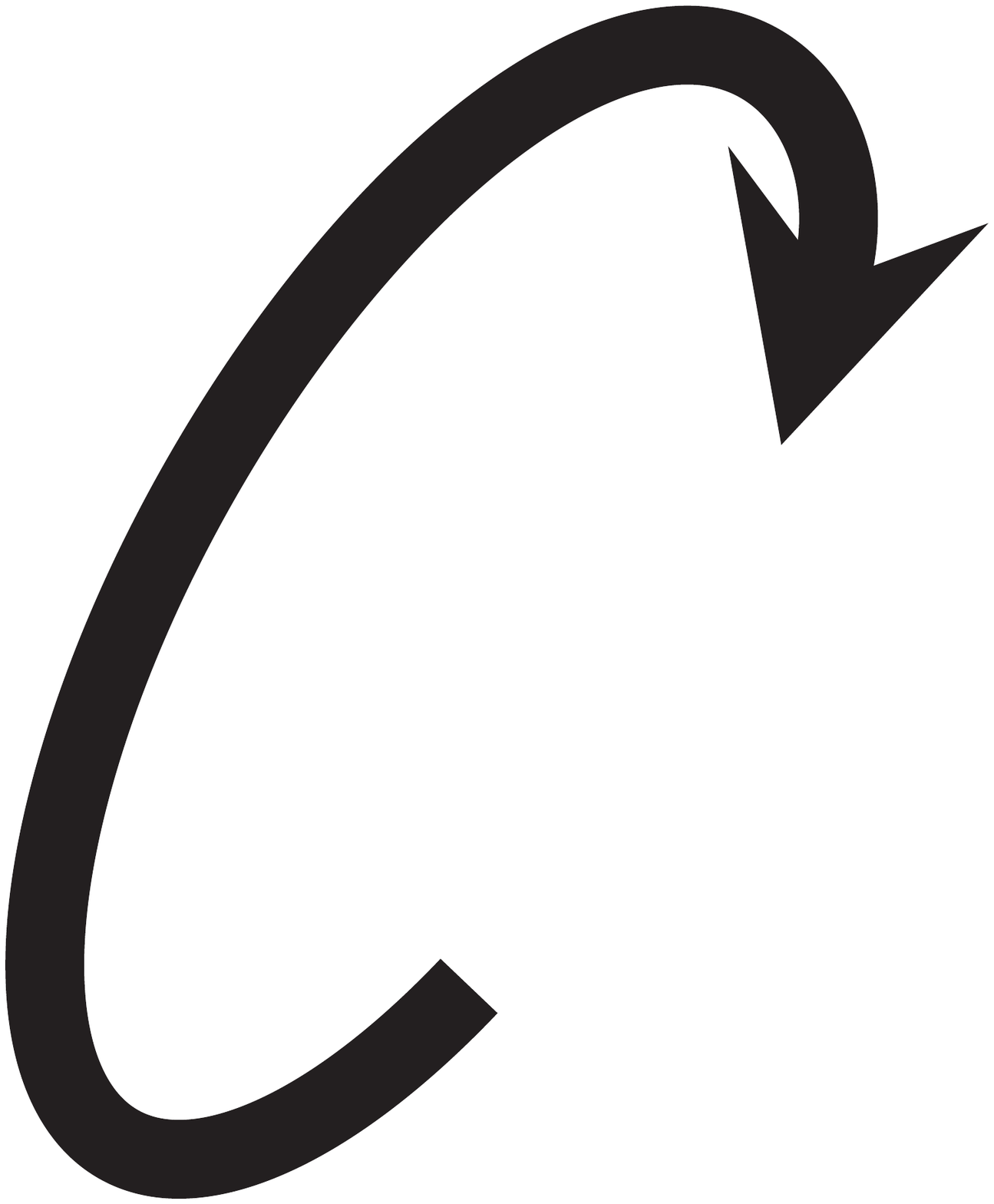}}
\put(50.3, 75){$\psi_I$}
\put(53.8, 74) {\Huge \bf .}
\put(40.5, 14) {$\psi_F$}
\put(44.7, 13.3) {\Huge \bf .}
\end{overpic}
\end{center}
\caption{
\label{fig:3}
(colour online) 
\textit{Quantum Zermelo navigation}. The `wind' ${\hat H}_0$ corresponds to an 
external field in the direction indicated by the red arrow; the associated flow lines are 
indicated by latitudinal circles. 
In this example, the wind 
is in the direction almost directly against the shortest path from $|\psi_I\rangle$ to 
$|\psi_F\rangle$. When the strength of the 
wind is close to zero, the path generated by the optimal Hamiltonian follows close 
to the geodesic curve. However, as the strength of the wind is increased, the 
optimal path is drifted away from the geodesic curve. Once the wind strength is 
sufficiently strong, it is optimal to go the other way around; the loss generated by the 
additional journey length is compensated by the fact that the head wind has turned 
into a tailwind. }
\end{figure}

To proceed with this let us rewrite (\ref{eq:4}) in the form 
\begin{eqnarray}
\re^{-{\rm i}{\hat H}_1(0) T} |\psi_I\rangle = \re^{{\rm i}{\hat H}_0 T} |\psi_F\rangle . 
\label{eq:5}
\end{eqnarray}
Although the transformation from (\ref{eq:4}) to (\ref{eq:5}) is in itself trivial, it sheds a different 
light on the problem at hand. As indicated above, if we transform to the moving frame generated 
by the wind such that the initial state remains stationary, then the objective becomes `hitting' a 
moving target $\re^{{\rm i}{\hat H}_0 t} |\psi_F\rangle$ as quickly as possible. Exactly how long it 
will take to achieve this task depends on how fast the motion towards the target can be, but for a 
fixed speed the shortest time $T$ achievable is attained by following the geodesic path on the 
state space joining $|\psi_I\rangle$ and $\re^{{\rm i}{\hat H}_0 T} |\psi_F\rangle$. On the other 
hand, the curve $|\psi(t)\rangle=\re^{-{\rm i}{\hat H}_1(0) t} |\psi_I\rangle$ joining $|\psi_I\rangle$ 
and $\re^{-{\rm i}{\hat H}_0 T} |\psi_F\rangle$ is a geodesic on the Fubini-Study manifold if and 
only if ${\hat H}_1(0)$ is horizontal, and this is indeed the situation we have in (\ref{eq:5}). To 
determine ${\hat H}_1(0)$ explicitly, let $\theta=\theta(T)$ be the angular distance between 
$|\psi_I\rangle$ and $\re^{{\rm i}{\hat H}_0 T} |\psi_F\rangle$ defined by the relation: 
\begin{eqnarray}
|\langle\psi_I|\re^{{\rm i}{\hat H}_0 T} |\psi_F\rangle| = \cos\half\theta .
\label{eq:6} 
\end{eqnarray}
Evidently, the angular distance $\theta(T)=\omega T$ is given by the speed $\omega$ of the 
evolution generated by 
${\hat H}_1(0)$ multiplied by the duration $T$ of the journey, but the evolution speed is fixed on 
account of the condition $2{\rm tr}({\hat H}_1^2)=1$ to $\omega=1$, so we deduce that 
\begin{eqnarray}
2 \arccos |\langle\psi_I|\re^{{\rm i}{\hat H}_0 T} |\psi_F\rangle| =  T .
\label{eq:7} 
\end{eqnarray}
Since ${\hat H}_0$, $|\psi_I\rangle$, and $|\psi_F\rangle$ are given, we see therefore that 
the smallest positive root of (\ref{eq:7}) uniquely determines $T$. Once $T$ is fixed, the 
optimal control Hamiltonian at time zero can be obtained explicitly by making use of the result 
in \cite{brody2}: 
\begin{eqnarray}
{\hat H}_1(0) = \frac{\ri 
\left( 
|\psi_I\rangle \langle\psi_F|\re^{-{\rm i}{\hat H}_0 T} - \re^{{\rm i}{\hat H}_0 T} 
|\psi_F\rangle \langle\psi_I| \right) }{2 \sin\left( \frac{1}{2}T\right)} . 
\label{eq:8} 
\end{eqnarray}
Note that without loss of generality we can adjust the overall phase such that $\langle\psi_I
|\re^{{\rm i}{\hat H}_0 T} |\psi_F\rangle$ is real. With this convention, which is assumed in 
(\ref{eq:8}), one can easily verify that this Hamiltonian is indeed horizontal. 

\begin{figure}[t]
\begin{center}
\includegraphics[scale=0.5]{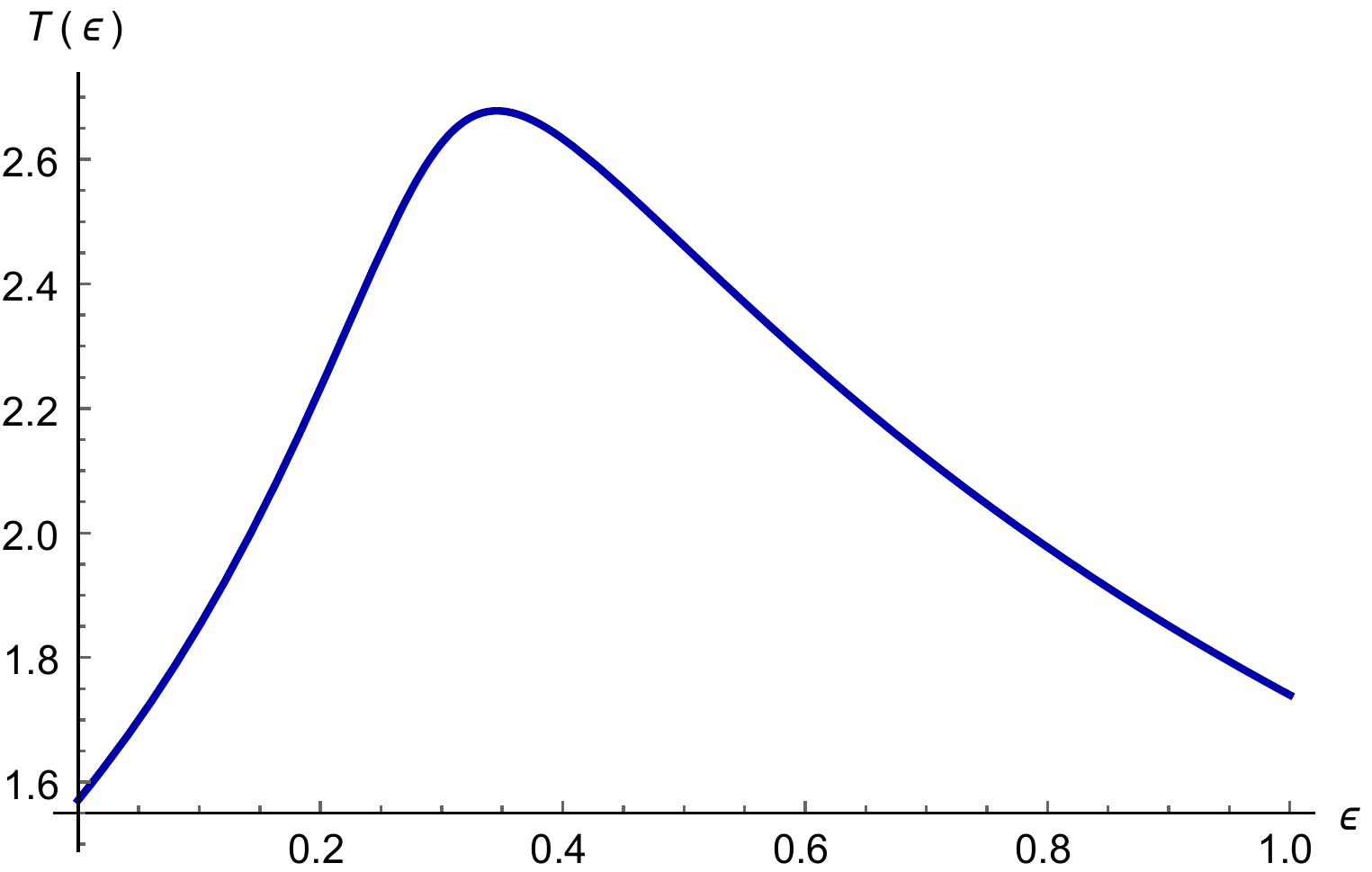}
\end{center}
\caption{
\label{fig:4}
(colour online) 
\textit{Journey time on the Bloch sphere}. 
The trajectories $|\psi(t)\rangle$ sketched in figure~\ref{fig:3} 
correspond to the wind Hamiltonian $\epsilon{\hat H}_0$ for $\epsilon = 0,~0.1,~ 
0.2,~ \ldots,~ 1.0$, for the choice of ${\hat H}_0$ sketched therein. Here, journey 
time $T(\epsilon)$ is plotted for a continuous range of $\epsilon$, showing that 
once the wind strength is sufficiently strong, by going the other way around, $T$ 
can be reduced by turning head wind into a tailwind. 
}
\end{figure}

This completes the derivation of the solution to the quantum-state Zermelo navigation 
problem: the optimal control Hamiltonian is given by (\ref{eq:2}), where ${\hat H}_1(0)$ 
is given by (\ref{eq:8}) and where the parameter $T$ is determined by (\ref{eq:7}). To 
gain further intuition on the behaviour of the solution, in figure~\ref{fig:3} we plot the 
trajectories of the quantum state generated by the Hamiltonian ${\hat H}(t)={\hat H}_0
+{\hat H}_1(t)$ in the case of a two-level system. In particular, we show the 
time-optimal paths for a range of wind strengths, indicating for instance that if the head 
wind is too strong, then the optimal strategy is to switch the direction of the manoeuvre 
to turn the head wind into a tailwind. To understand the minimum journey times 
associated with different wind strengths we have also considered a 
one-parameter family of winds $\epsilon{\hat H}_0$, corresponding to the setup shown 
in figure~\ref{fig:3}, and determined the optimal time $T(\epsilon)$ for a range of 
$\epsilon$; the result is shown in figure~\ref{fig:4}. 

In summary, 
we have been able to deduce in closed form the solution to the quantum Zermelo 
problem in complete generality. Since 
the result obtained takes a rather simple form, we expect that a practical 
implementation is feasible in a number of realistic situations. Of course, 
for some applications the controllable degrees of freedom in the Hamiltonian can be 
limited, in which case additional constraints will have to be imposed in finding the 
optimal control---some steps towards this direction in the context of unitary gates 
have been initiated in \cite{stepney3}. In fact, further progress can again be made by 
switching to the interaction picture: If we write $\{f_k({\hat H}_1)=0\}_{k=1,\ldots,N}$ 
for the $N$ constraints on the control Hamiltonian, then in the moving frame the 
constraints will take for each $k$ the form 
$f_k(\re^{-{\rm i}{\hat H}_0 t}{\hat h}\re^{{\rm i}{\hat H}_0 t})=0$, where 
${\hat h}=\re^{{\rm i}{\hat H}_0 t}{\hat H}_1\re^{-{\rm i}{\hat H}_0 t}$ is the generator 
of the dynamics $\ri\partial_t|\psi\rangle={\hat h}|\psi\rangle$. The analysis of 
\cite{hosoya}, for instance, can then be applied with essentially just one modification, 
namely, that the target state is moving. 
The initial control Hamiltonian can then be identified by 
adapting the strategy proposed in \cite{lloyd}. 

It is worth noting that 
the use of the interaction picture, or equivalently the switch to the moving 
frame, which allowed us to simplify the problem considerably, has implications beyond 
the quantum navigation problem considered here. In this connection we remark that if we endow 
the quantum state space with the metric of the form (\ref{eq:1}), then the resulting space 
is known as a Randers space (cf. \cite{gibbons2}), which is an example of a Finsler 
space (for $|w|\to1$ the space reduces to a Kropina space). 
In the literature of Finsler geometry it is known that every Zermelo navigation 
problem on a Riemannian manifold can be solved by finding geodesics of the 
corresponding Randers space \cite{robles0}. The classification of spaces in terms of 
their curvatures is of particular importance in geometry, and in the case of Randers spaces 
the identification of constant curvature metrics has been obtained by solving the 
associated navigation problem when the wind is a conformal vector field 
with constant scale factor \cite{robles} (the field generated by a 
unitary motion belongs to this class, with vanishing scale factor). However, and with 
hindsight, the main theorem of \cite{robles} can be proven in essentially two 
lines by adopting the interaction picture, and this shows that in the present context a 
physical intuition can  offer insights into a purely mathematical question. We conclude 
by remarking that an interesting generalisation of the present problem is to consider the 
case under which the external influences contain noise. In such a situation, the idea of 
reaching a target pure state is no longer tenable, whereas reaching a particular mixed 
state might be feasible. Although there appears to be a surprisingly limited amount of work in 
the stochastic extensions of the Zermelo navigation problem (cf. \cite{ECL}), extensions 
into the mixed-state domain are desirable for the designing and 
understanding of more robust controlled quantum dynamics.

\vspace{0.2cm} 

\begin{acknowledgments}
We thank Eva-Maria Graefe, Benjamin Russell, and Dmitry Savin for discussion and 
comments, and Martin Meier for support in producing figure~\ref{fig:2}. 
\end{acknowledgments}

\onecolumngrid

\newpage 


\section*{Supplementary Information:  Derivation of the universal quantum speed limit}  
\noindent 
The purpose of this appendix is to offer a proof of the following claim in the paper: 
\emph{The squared speed of the evolution of a quantum state generated by 
a Hamiltonian $\hat{H}$ is bounded above by twice the Hilbert-Schmidt norm 
${\rm tr}(\hat{H}^2)$ of the Hamiltonian, and the bound is attained if $\hat{H}$ 
is horizontal.} 

We begin by establishing some properties of the space of horizontal Hamiltonians. 
Recall that a Hamiltonian $\hat{H}$ is horizontal with respect to some state 
$|\psi\rangle$ if and only if ${\rm tr}(\hat{H} \hat{H}_{|\psi \rangle}) = 0$ for all 
Hamiltonians $\hat{H}_{|\psi\rangle}$ that leave $|\psi \rangle$ invariant. 
For simplicity, we fix the state $|\psi \rangle$ to be the one whose homogeneous 
coordinates are given by $(1, 0, \ldots, 0)^T \in \mathds{C}^{n+1}$. It should be 
stressed, however, that the conclusions of the discussion below remain valid for any 
choice of $|\psi\rangle$, owing to the homogeneous nature of the state space 
${\mathds C}{\mathbb P}^{n}$. Let us write $V_{|\psi\rangle}$ for the subgroup of 
SU$(n+1)$ that leaves $|\psi \rangle$ invariant and $\mathfrak{v}_{|\psi \rangle}$ for 
the set of the corresponding generators (i.e. its Lie algebra). Then $V_{|\psi\rangle}$ 
consists of all unit determinant matrices of the form
\begin{eqnarray}
\left(\begin{array}{cc} \re^{{\rm i} \theta} & \boldsymbol{0}^T \\ \boldsymbol{0} & U_n 
\end{array}\right), \qquad U_n \in U(n), \quad \boldsymbol{0}^T=(0,0,\ldots,0), 
\nonumber 
\end{eqnarray}
and $\mathfrak{v}_{|\psi\rangle}$ consists of all trace-free matrices of the form 
\begin{eqnarray}
\left(\begin{array}{cc} \lambda & \boldsymbol{0}^T \\ \boldsymbol{0} & {\hat B} \end{array} 
\right), \qquad  \lambda 
\mbox{ imaginary and }  {\hat B} \in \mathfrak{u}(n). \nonumber 
\end{eqnarray}
The orthogonal complement $\mathfrak{v}_{|\psi\rangle}^\perp$ of this space with respect 
to the Hilbert-Schmidt norm coincides, up to a factor of $\ri$, with the space of horizontal Hamiltonians. 
Specifically, the set $\mathfrak{v}_{|\psi\rangle}^\perp$ can be found by requiring that 
\begin{eqnarray}
{\rm tr} \left( \left(\begin{array}{cc} \lambda & \boldsymbol{0}^T \\ \boldsymbol{0} & {\hat B} 
\end{array} \right)  
  \left(\begin{array}{cc} \mu & -\bar{\boldsymbol{z}}^T  \\ \boldsymbol{z} & {\hat C} 
  \end{array} \right) \right) 
  = \lambda \mu + {\rm tr}({\hat B}{\hat C}) = 0 
\end{eqnarray}
for all choices of $\lambda$ and ${\hat B}$ with $\lambda + {\rm tr}({\hat B}) = 0$. Here we have 
parameterised the full Lie algebra $\mathfrak{su}(n+1)$ using an imaginary $\mu$, a vector 
$\boldsymbol{z} \in \mathds{C}^{n}$ and ${\hat C} \in \mathfrak{u}(n)$ with 
$\mu + {\rm tr}({\hat C}) = 0$. It follows 
that $\mathfrak{v}_{|\psi\rangle}^\perp$ consists of all matrices of the form
\begin{eqnarray}
    \left(\begin{array}{cc} 0 & -\bar{\boldsymbol{z}}^T  \\ \boldsymbol{z} & \mathbb{0} 
    \end{array} \right), \nonumber 
\end{eqnarray}
where $\boldsymbol{z} \in \mathds{C}^{n}$. As a consistency check, we note that this space 
has $2n$ real 
dimensions, which of course is the same as the dimensionality of the state space 
$\mathds{C}{\mathbb P}^{n}$. It follows that if $\hat{H}$ is a horizontal Hamiltonian, then
\begin{eqnarray}
2 \, {\rm tr}( \hat{H}^2) = - 2\, {\rm tr}  \left(\left(\begin{array}{cc} 0 & -\bar{\boldsymbol{z}}^T 
\\  \boldsymbol{z} & \mathbb{0} 
\end{array} \right)^2\right) = 4 |\boldsymbol{z}|^2. 
\label{app_compare}
\end{eqnarray}
On the other hand, if $|\psi\rangle$ evolves according to the Schr\"odinger equation, then 
the squared speed (in the Fubini-Study metric) of its evolution is given, on account of the 
Anandan-Aharonov relation, by four times the variance of $\hat{H}$:
\begin{eqnarray}
  4 \Delta H^2 = 4  ( \langle \psi | \hat{H}^2 | \psi \rangle - \langle \psi | \hat{H}| \psi 
  \rangle^2) = 4 |\boldsymbol{z}|^2. \label{app_compare2}
\end{eqnarray}
This establishes part of our claim: Namely, by comparing \eqref{app_compare} and 
\eqref{app_compare2} we see that the square of the evolution speed equals twice the 
Hilbert-Schmidt norm ${\rm tr}(\hat{H}^2)$ if $\hat{H}$ is horizontal. To complete the proof, note 
that any Hamiltonian $\hat{H}$ can be decomposed into vertical and horizontal parts, 
i.e. it can be written in the form
\begin{eqnarray}
  \hat{ H}= \ri   \left(\begin{array}{cc} \lambda & \boldsymbol{0}^T \\ \boldsymbol{0} & {\hat B} 
  \end{array} \right) + 
  \ri \left(\begin{array}{cc} 0 & -\bar{\boldsymbol{z}}^T \\ \boldsymbol{z} & \mathbb{0} 
  \end{array} \right).
\end{eqnarray}
Then
\begin{eqnarray}
 2\,{\rm tr}(\hat{H}^2)= 4|\boldsymbol{z}|^2 - 2 {\rm tr}\left( \left(\begin{array}{cc} \lambda 
 & \boldsymbol{0}^T \\ \boldsymbol{0} & {\hat B} 
 \end{array} \right)^2\right),
\end{eqnarray}
but the trace on the right hand side is strictly negative if the vertical part of $\hat{H}$ is 
nonzero. Therefore, the squared speed of evolution on state space, given by $4
|\boldsymbol{z}|^2$, is strictly smaller than $2\,{\rm tr}(\hat{H}^2)$ if the Hamiltonian contains 
a vertical component. This completes the proof of the claim.

\end{document}